\documentclass[submission,copyright,creativecommons]{eptcs}
\providecommand{\event}{PLACES 2012} 
\usepackage{breakurl}             

\usepackage[english]{babel}
\usepackage{amsmath}
\usepackage{amsfonts}
\usepackage{amsthm}
\usepackage{semantic}

\usepackage{xcolor}

\newcommand{\coml}{ContextML}

\newcommand{\mylet}{\text{\textbf{let}}\;}
\newcommand{\myin}{\;\text{\textbf{in}}\;}

\newcommand{\mywith}{\text{\textbf{with}}}

\newcommand{\myfun}{\mathbf{\lambda}} 
\newcommand{\mylayernames}{\mathsf{LayerNames}}
\newcommand{\myif}{\text{\textbf{if}}\;}
\newcommand{\mythen}{\;\text{\textbf{then}}\;}
\newcommand{\myelse}{\;\text{\textbf{else}}\;}

\theoremstyle{plain}
\newtheorem{theorem}{Theorem}[section] 
\newtheorem{lemma}[theorem]{Lemma}

\theoremstyle{definition}

\theoremstyle{remark}

\title{Typing Context-Dependent Behavioural Variations}
\def\titlerunning{Typing Context-Dependent Behavioural Variations}
\author{Pierpaolo Degano
\quad \quad
Gian-Luigi Ferrari
\quad \quad
Letterio Galletta
\quad \quad
Gianluca Mezzetti
\email{degano@di.unipi.it \quad \quad giangi@di.unipi.it \quad \quad galletta@di.unipi.it \quad \quad mezzetti@di.unipi.it}
\institute{{Dipartimento di Informatica, Universit\`a di Pisa, Pisa, Italy}}
}

\def\authorrunning{Degano, Ferrari, Galletta and Mezzetti}

\begin{document}

\maketitle

\begin{abstract}
Context Oriented Programming (COP) concerns the ability of programs to adapt to changes in their running environment. 
A number of programming languages endowed with COP constructs and features have been developed.
However, some foundational issues remain unclear.
This paper proposes adopting static analysis techniques to reason on and predict how programs adapt their behaviour. 
We introduce a core functional language, ContextML, equipped with COP primitives for manipulating contexts and for programming behavioural variations.
In particular, we specify the dispatching mechanism, used to select the program fragments to be executed in the current active context.
Besides the dynamic semantics we present an annotated type system.
It guarantees that the well-typed programs adapt to any context, i.e. the dispatching mechanism always succeeds at run-time.
\end{abstract}

\section{Introduction}

Computers increasingly pervade our everyday life, pushed by the great improvements of hardware systems and by the growing usage of digital information. 

On the one side, computer devices surround people in different shapes and sizes in a highly distributed manner.
Devices are often interconnected, can interact each others and share resources.
Programs themselves are resources being invokable remote services or downloadable code.

On the other side, processing the great quantity of information generated and consumed by devices leads to a paradigm, where most of the computation and of the storage is demanded to specific, powerful remote entities.
This is the case of Cloud or Grid systems, made of heterogeneous nodes, possibly distributed in a wide-area.

This new setting puts software system into a highly-dynamic environment where services, resources and hardware components appear, mutate, move and disappear.
This calls for a a great shift of programming paradigm.
In particular, there is a growing interest in the design and development of applications that are aware of their working environment and are \emph{adaptive}, i.e.~capable to adapt to different situations.

Context-Oriented Programming (COP)\cite{Costanza08programming} has been recently proposed to tackle such an issues. 
COP makes available language primitives to express context-dependent behaviour (called \emph{behavioural variation}) in a modular fashion.
Behavioural variations are chunks of behaviour that modify the execution of a computational system depending on the current working environment.
Layers are the linguistic mechanism that enable the programmer to group variations.
Layers can be activated in arbitrary places of the code with appropriate primitives.
Each layer activation has its own scope; activated layers are piled up into a stack that is called \emph{context}.
Therefore, the actual behaviour of a COP program is carried out by a \emph{dispatching} procedure that selects the program fragments to be executed depending on the context.

Most of the research efforts in this field are focused on the concrete implementation details and only a few papers~\cite{hirschfeld2011contextfj,Clarke2009} investigated basic foundational issues.
We briefly discuss them in Section~3.

Our work aims at contributing to the foundations of COP programming languages.
We introduce a core calculus (\coml{}) with a precise semantic description of the adaptivity constructs.

To illustrate the novel features of \coml{} we resort to a running example.
Consider a program embedded in a mobile device, the 
behaviour of which depends on the active profile of its battery.
We assume that the battery level has two profiles, the power saving mode and the performance one.
These profiles are represented at code level as two different layers: \verb|PowerSavingMode| and \verb|PerformanceMode|.
The function \verb|getBatteryProfile|, described below, queries the sensor (\verb|batSensor|) and returns the layer describing the current active profile depending on a \verb|threshold| value:
{\small
\begin{align*}
\lambda_{\mathtt{getBatteryProfile}} () \Rightarrow
     &\mathtt{\myif (batSensor() > threshold) \mythen} \\
     	&\qquad\mathtt{PerformanceMode} \\
     &\myelse \\ 
     	&\qquad\mathtt{PowerSavingMode}
\end{align*}
}
Layers are expressible values, hence they can be produced as results of function calls.
The construct $\mywith(e_{1})\myin e_{2}$ activates the layer obtained evaluating $e_{1}$ and delimits its activation scope to the inner expression $e_{2}$.
For instance, in the code below, the layer obtained as result of the call $\mathtt{getBatteryProfile()}$ is active throughout the execution of the inner expression.
{\small
\begin{align*}
\mywith&(\mathtt{getBatteryProfile()}) \myin \\
  &\mathtt{PowerSavingMode.\:doSomeThing()},\\
  &\mathtt{PerformanceMode.\:doSomeThingElse()}
\end{align*}
}
In the inner expression, the \emph{layered expression} is defined by cases that specify the context-dependent behavioural variation considered.

Note that if the programmer neglects a case for the profile of the battery, e.g.\ \verb|OnDemandMode|, then the program throws a run-time error being unable to adapt to the context.
We propose to detect these undesired behaviour by adopting static analysis techniques.
To this aim we extend the ML type system in order to guarantee that well-typed programs are always capable to react to their changing environment, i.e.\ the dispatching procedure always succeeds at run-time.

\section{\coml : a context-oriented ML core}
\coml\ is a purely functional fragment of ML extended with COP primitives. 
In \coml\ the context is explicit and is part of the run-time environment.
The language is endowed with primitives to manipulate the context and to specify behavioural variation of expressions, depending on the context in which the program is evaluated.
The structural operational semantics and the type system of \coml{} follow.
\paragraph{Dynamic Semantics}
Let $\mathbb{N}$ be the set of naturals, $\mathsf{Ide}$ a set of identifiers, $\mathsf{LayerNames}$ a set of layer names, then the syntax of \coml\ is defined by the following grammar:
{\small
\begin{align*}
n &\in \mathbb{N} \qquad x,f \in \mathsf{Ide} \qquad L \in  \mylayernames \\
v,v_{1},v' &::= n \mid L \mid \myfun_{f} \; x \; \Rightarrow e \\
e,e_{1},e'   &::= v \mid x \mid e_{1} e_{2} \mid \mylet x = e_{1} \myin e_{2} \mid e_{1} \;\text{\textbf{op}} \; e_{2} \mid 
 \myif e_{0}  \mythen e_{1} \myelse e_{2} \mid \mywith(e_{1}) \myin e_{2} \mid lexp \\
lexp & ::= L.e \mid L.e , lexp
\end{align*}
}
The novelties with respect to ML are layers as expressible values; the $\mywith$ construct for activating layers; and the layered expressions ($lexp$).
Recall that a context $C$ is a stack of active layers. We denote with $L::C$ the pushing of layer $L$ on $C$ and with $[L_{1},\dots,L_{n}]$ a context with $n$ elements whose top is $L_{1}$.

The semantics is only defined for closed expression and is characterised by judgements having the form $C \vdash e \rightarrow e'$ meaning that in context $C$ the closed expression $e$ reduces to $e'$ in one evaluation step.
In \figurename~\ref{fig:comlsem} we only show the semantic rules (with$_{1}$),(with$_{2}$),(with$_{3}$),(lexp) that deal with the new constructs, the others are inherited from the standard ML.
\begin{figure}
\centering
{\def\spskip{10pt}\small
\begin{tabular}{cc}
\inference[with$_{1}$]{L::C \vdash e \rightarrow e'}{C \vdash \mywith(L) \myin e \rightarrow \mywith(L) \myin e'}
&
\inference[with$_{2}$]{}{C \vdash \mywith(L) \myin v \rightarrow v}
\\[\spskip]
\inference[with$_{3}$]{C \vdash e \rightarrow e'}{C \vdash \mywith(e) \myin e_{1} \rightarrow  \mywith(e') \myin e_{1} }
&
\inference[lexp]{\exists k.k = \min \{j \mid \exists v.L'_{j} = L_{v} \} \land L'_{k} = L_{i}}{[L'_{1},\dots,L'_{m}] \vdash  L_{1}.e_{1}, \dots , L_{n}.e_{n} \rightarrow e_{i}}
\end{tabular}
}
\caption{\coml{} semantics.}
\label{fig:comlsem}
\end{figure}
We briefly comment on the new ones only.
Rules for  $\mywith(e_{1}) \myin e_{2}$ evaluate $e_{2}$ in the context extended by the layer obtained evaluating $e_{1}$.
When a layered expression $L_{1}.e_{1}, \dots , L_{n}.e_{n}$ has to be evaluated (rule \emph{lexp}), the context of evaluation is inspected top-down (dispatch mechanism). 
When a layer in the context matches one of the $L_{i}$, the corresponding expression $e_{i}$ is evaluated. If no layer matches then the computation gets stuck.

\paragraph{Type system}
We introduce a monomorphic type system for \coml\ which ensures that the dispatch mechanism always succeeds at run-time for well-typed expressions.

Our type system is characterised by typing judgements of the form $\langle \Gamma;C \rangle \vdash e : \tau$. 
This means that in ``in the type environment $\Gamma$ and in the context $C$ expression $e$ has type $\tau$''.

Types are integers, layers and functions.
\[
\tau,\tau_{1},\tau'::= int \mid ly_{\phi} \mid \tau_{1} \xrightarrow{\psi} \tau_{2} \qquad
\phi,\psi \in \wp(\mylayernames) 
\]
We annotate types with sets of layer names $\phi,\psi$ for analysis reason.
In $ly_{\phi}$, $\phi$ over-approximates the layers that an expression can be reduced to at run-time.
In $\tau_{1} \xrightarrow{\psi} \tau_{2}$, $\psi$ over-approximates the layers that must be active in the context during the application of the function (precondition of the function).

Back to our example, the type of the function \verb|getBatteryProfile| will be the following: 
\[
\mathtt{unit} \xrightarrow{\emptyset} ly_{\{\mathtt{PowerSavingMode},\:\mathtt{PerformanceMode}\}}.
\]
The intuition is that the function returns a layer in the set $\{ \mathtt{PowerSavingMode},
\mathtt{PerformanceMode}\}$. 
The function has no preconditions, i.e.\ it can be applied in any context.

\begin{figure}[t]
\centering
{\def\spskip{10pt}\small
\begin{tabular}{cc}
\inference[(Sint)]{}{int \leq int}
&
\inference[(Sly)]{\phi \subseteq \phi'}{ly_{\phi} \leq ly_{\phi'}}
\\[\spskip]
\multicolumn{2}{c}{\inference[(Sfun)]{\tau_{1}' \leq \tau_{1} & \tau_{2} \leq \tau_{2}' & \psi \subseteq \psi'}{\tau_{1} \xrightarrow{\psi} \tau_{2} \leq \tau_{1}' \xrightarrow{\psi'} \tau_{2}'}}
\\[18pt]
\inference[(Tint)]{}{\langle \Gamma;C \rangle \vdash n : int}
&
\inference[(Tly)]{}{\langle \Gamma;C \rangle \vdash L: ly\{L\}} 
\\[\spskip]
\inference[(Tsub)]{\langle \Gamma ; C \rangle \vdash e : \tau' & \tau' \leq \tau}{\langle \Gamma ; C \rangle \vdash e : \tau}
&
\inference[(TVar)]{\Gamma(x)=\tau & \text{if $x \in dom(\Gamma)$}}{\langle \Gamma;C \rangle \vdash x : \tau} 
\\[\spskip]
\multicolumn{2}{c}{\inference[(Tfun)]{\langle \Gamma,x:\tau_{1},f:\tau_{1} \xrightarrow{\left | C' \right |} \tau_{2}; C' \rangle \vdash e:\tau_{2}}{\langle \Gamma; C \rangle \vdash \myfun_{f} \; x \Rightarrow e : \tau_{1} \xrightarrow{\left | C' \right |} \tau_{2}} }
\\[20pt]
\multicolumn{2}{c}{\inference[(Top)]{\langle \Gamma;C \rangle \vdash e_{1} : int  & \langle \Gamma;C \rangle \vdash e_{2} : int }{\langle \Gamma;C \rangle \vdash e_{1} \;\textbf{op}\; e_{2} : int } }
\\[\spskip]
\multicolumn{2}{c}{\inference[(Tlet)]{\langle \Gamma; C \rangle \vdash e_{1}:\tau_{1} & \langle \Gamma, x:\tau_{1} , C \rangle \vdash e_{2} : \tau_{2}}{\langle \Gamma; C \rangle \vdash \mylet x = e_{1} \myin e_{2}:\tau_{2}}}
\\[\spskip]
\multicolumn{2}{c}{\inference[(Tif)]{\langle \Gamma; C \rangle \vdash e_{0}:int & \langle \Gamma; C \rangle \vdash e_{1}:\tau & \langle \Gamma; C \rangle \vdash e_{2}:\tau} {\langle \Gamma; C \rangle \vdash \myif \; e_{0} \; \mythen e_{1} \myelse e_{2} : \tau} }
\\[\spskip]
\multicolumn{2}{c}{\inference[(Twith)]{\langle \Gamma;C \rangle \vdash: e_{1}: ly_{\phi} & \forall L' \in \phi . \langle \Gamma;L'::C \rangle \vdash e_{2}: \tau}{\langle \Gamma; C \rangle \vdash \mywith (e_{1}) \myin e_{2} : \tau}}
\\[\spskip]
\multicolumn{2}{c}{\inference[(Tlexp)]{\forall i. \langle \Gamma;C \rangle \vdash e_{i}: \tau  & L_{1} \in |C| \vee \dots \vee L_{n} \in |C| }{\langle \Gamma; C \rangle \vdash L_{1}.e_{1} , \dots , L_{n}.e_{n}: \tau}}
\\[\spskip]
\multicolumn{2}{c}{\inference[(Tapp)]{\langle \Gamma;C \rangle \vdash e_{1}: \tau_{1} \xrightarrow{\phi} \tau_{2} &  \langle \Gamma;C \rangle \vdash e_{2} : \tau_{1} & \phi \subseteq |C| }{\langle \Gamma; C \rangle \vdash e_{1}e_{2} : \tau_{2}}}
\end{tabular}
}
\caption{\coml\ type system}
\label{fig:type}
\end{figure}

Our typing rules are in \figurename~\ref{fig:type}.
 Since types are annotated, the type system contains rules dealing with the subtyping. The rules (Sint), (Sly), (Sfun) have judgements of the form $\tau_1 \leq \tau_2$ ($\tau_1$ is a subtype
of $\tau_2$). 
Furthermore, we assume that annotations are ordered by set-inclusion and that $|C|$ is the set of active layers in a context $C$.

By rule (Sly) a layer type $ty_{\phi}$ is a subtype $ty_{\phi'}$ if and only if the annotation $\phi$ is a subset of $\phi'$.
Rule (Sfun) is subtyping rule for functional types. As usual $\tau_1 \xrightarrow{\psi} \tau_2$ is contravariant in $\tau_1$ but covariant
in $\phi$ and $\tau_2$.

Rule (Tly) asserts that the type of a layer $L$ is $ly$ annotated with the singleton set $\{L\}$.
In rule (Tfun) we guess a type for the bound variable, for the function $f$ and determine the type of the body
under these additional assumptions and in a guessed context $C'$. 
Implicitly, we require that the guess of a type for $f$ matches that of the resulting function. Additionally we require that the resulting type is annotated with a precondition that includes the layers in $C'$.

Rule (Twith) establishes that an expression $\mywith$ has type $\tau$, provided the type for $e_1$ is $ly_{\phi}$ 
(recall that $\phi$ is a set of layers) and $e_2$ has type $\tau$ 
in the context $C$ extended by the layers in $\phi$.
By (Tlexp) the type of a layered expression is $\tau$, provided that each sub-expression $e_i$ has type $\tau$ and that at least one among
the layers $L_1, \ldots L_n$ is active in the context $C$. 
This requirement is the key to ensure that the dispatch mechanism always succeeds at run-time.
Notably, when evaluating a layered expression one of the mentioned layers will be active in the current context.

Back to our example, the expression provided is well-typed as witnessed in \figurename~\ref{fig:deriv}.
The type of \verb|getBatteryProfile| ensures that one of the two layers \verb|PowerSavingMode| or \verb|PerformanceMode| is returned. One among them is required to be active so to evaluate the layered expression. 
Hence, the (Twith) rule can guarantee that the whole expression is never stuck at run-time.

\begin{figure}[t]
\scriptsize
\hspace{-5pt}\inference{
	\langle \Gamma;C \rangle \vdash: \mathtt{getBatteryProfile()}: ly_{\phi} &
	\hspace{-80pt}\inference[(Tlexp)]
	{
		 \langle \Gamma;C' \rangle \vdash \mathtt{doSomeThing()}: \tau  & \langle \Gamma;C' \rangle \vdash \mathtt{doSomeThingElse()}: \tau & \\
		 {\begin{array}{c}\mathtt{PowerSavingMode} \in |C'| \vee  \mathtt{PerformanceMode} \in |C'|\end{array}}
	}
	{
		\langle \Gamma; C' \rangle \vdash \hspace{-10pt}\begin{array}{c}\mathtt{PowerSavingMode.doSomeThing(),}\\\mathtt{PerformanceMode.doSomeThingElse()}\end{array}\hspace{-5pt}: \tau
	}
	&
	\hspace{-20pt}\inference[(Tlexp)]{\cdots}{\langle \Gamma; C'' \rangle \vdash \dots}
}
{
	\langle \Gamma; C \rangle \vdash \mywith (\mathtt{getBatteryProfile()}) \myin \begin{array}{c}\mathtt{PowerSavingMode.doSomeThing(),}\\\mathtt{PerformanceMode.doSomeThingElse()}\end{array} : \tau
}\\
\caption{\small The typing derivation of the running example. We assume $\mathtt{doSomeThing,doSomeThingElse}:\mathtt{unit}\rightarrow\tau$. We denote $\phi = \{\mathtt{PowerSavingMode,PerformanceMode}\}$;$C' =  \mathtt{PowerSavingMode}::C$;$C'' =  \mathtt{PerformanceMode}::C$. The last rule used is (Twith).
}
\label{fig:deriv}
\end{figure}

Rule (Tapp) is almost standard and reveals the mechanism of function precondition. The application gets a type  if only if the layers in the precondition $\phi$ are active in the current context $C$.
To better explain how preconditions work, consider the example in \figurename~\ref{fig:fundev}. 
There the function $\myfun_{f} \; x \Rightarrow L_{1}. 0$ is shown having type $int \xrightarrow{\{L_{1}\}} int$. 
This means that $L_{1}$ must be active in the context of activation of the function.

The remaining rules are standard and we do not comment on them for brevity.

\begin{figure}[t]

{\footnotesize
\centering
\inference[]{
	\inference{
		\inference{\langle \Gamma,x: \tau,f:\tau \rightarrow \tau ; C' \rangle \vdash 0 : \tau & L_{1} \in C'}
	{\langle \Gamma,x: \tau,f:\tau \xrightarrow{\left | C' \right |} \tau ; C' \rangle \vdash L_{1}.0: \tau }
	}{\langle \Gamma ; C \rangle \vdash \myfun_{f} \; x \Rightarrow L_{1}. 0 : \tau \xrightarrow{\left | C' \right |} \tau}
	&
	\inference{
	\hspace{-31pt}\langle \Gamma,g: \tau \xrightarrow{\left | C' \right |} \tau ; C \rangle \vdash g: \tau \rightarrow \tau 
	\\
	\langle \Gamma,g: \tau \xrightarrow{\left | C' \right |} \tau ; C \rangle \vdash 3: \tau
	&
	\left | C' \right | \subseteq \left | C \right |
	}{\langle \Gamma,g: \tau \xrightarrow{\left | C' \right |} \tau; C \rangle \vdash g\,3 : \tau}
}{\langle \Gamma ; C \rangle \vdash \mylet g = \myfun_{f} \; x \Rightarrow L_{1}. 0 \myin g\,3 : \tau}
}
\caption{Derivation of a function with precondition. We assume that $C' = [L_{1}]$, $L_{1}$ is active in $C$ and, for typesetting convenience, we also denote $\tau = int$.}
\label{fig:fundev}
\end{figure}

Our type system guarantees not only that functional types are correctly used, but also that the evaluation of a layered expression never gets stuck. The following lemmata prove that our type system is sound with respect to the operational semantics:
\begin{lemma}[Progress]
Let $e$ be a closed expression such that for some $C$ $\langle \Gamma;C \rangle \vdash e : \tau$. Then either $e$ is a value or $C \vdash e \rightarrow e'$.
\end{lemma}
\begin{lemma}[Subject reduction]
Let $e$ be a closed expression, if $\langle \Gamma;C \rangle \vdash e : \tau$ and $C \vdash e \rightarrow e'$ then $\langle \Gamma, C \rangle \vdash e' : \tau$
\end{lemma}

\section{Discussion}

\paragraph{Related work}
Several COP programming languages have been proposed (see e.g.\emph{ContextL}~\cite{Costanza05languageconstructs} and \emph{ContextJ}~\cite{appeltauer2011contextj}).
Usually COP features are introduced within the object oriented paradigm so providing behavioural variations at object level. 

Most of the research efforts have mainly tackled implementation issues.
To the best of our knowledge only few papers provide a precise semantic description. 

In \cite{hirschfeld2011contextfj} an extension of \emph{Featherweight Java} \cite{IgarashiPW01} has been proposed.
This calculus includes \emph{layers (de)activation}, but layers are not expressible values.
Furthermore, a static type system ensures that there exists a binding for each dispatched method call.
This fact is based on the strong assumption that layers do not introduce new methods but only refine existent ones.
Our type system relaxes this assumption.

Our approach is much similar to the one of Clarke et al.~\cite{Clarke2009} and the main difference is that we consider a functional language while~\cite{Clarke2009} considers an object oriented language, Featherweight Java.

\paragraph{Conclusions and further work}
We started investigating the foundational issues of the COP paradigm with a calculus endowing COP primitives.
We have defined a dynamic semantics that formalises the operational mechanisms behind these constructs.
A distinguished element of our semantics is the dispatching procedure, that selects the behavioural variation depending on the active context.
We have also specified a type system guaranteeing that the dispatching mechanism always succeeds at run-time for well-typed expressions.

In our current proposals, activation of layers is driven according to the flow of program execution.
A more general approach would instead consider the asynchronous evolution of the environment.
We plan to investigate this issue and to formalise event-driven changes of context.

We also intend to refine the type system by introducing effects to represent an over-approximation of the evolution pattern of context shape.
In doing that, techniques similar to session-types~\cite{honda1998} might suggest us useful mechanisms.
Types and effects will enhance our static analysis of programs, following the lines of~\cite{Bartoletti1,Bartoletti2}. 
In particular we would like to accept or reject programs at compile time, also depending on non-functional requirements on the context evolution, e.g.\ when security policies are to be enforced upon context usages (for a recent proposal see \cite{coordination2012}).

\paragraph{Acknowledgement}
The authors would like to thank the anonymous referees for their comments that pointed us an inaccuracy, and guided us to improve the quality of our paper.

This work has been partially supported by IST-FP7-FET open-IP project ASCENS and Regione Autonoma Sardegna, L.R. 7/2007, project TESLA.

\bibliographystyle{eptcs}
\bibliography{bib}

\end{document}